\documentclass[twocolumn,aps,superscriptaddress,amsmath,amssymb]{revtex4-1}
\usepackage{amsfonts}
\usepackage{bbm}
\usepackage{tipa}

\usepackage{graphicx}
\usepackage{dcolumn}
\usepackage{bm}
\begin{document}

\title{Topologically nontrivial and trivial zero modes in chiral molecules}

\author{Xiao-Feng Chen}
\affiliation{Hunan Key Laboratory for Super-microstructure and Ultrafast Process, School of Physics and Electronics, Central South University, Changsha 410083, China}
\affiliation{School of Physical Science and Technology, Lanzhou University, Lanzhou 730000, China}

\author{Wenchen Luo}
\affiliation{Hunan Key Laboratory for Super-microstructure and Ultrafast Process, School of Physics and Electronics, Central South University, Changsha 410083, China}

\author{Tie-Feng Fang}
\affiliation{School of Sciences, Nantong University, Nantong 226019, China}

\author{Yossi Paltiel}
\affiliation{Applied Physics Department and the Centre for Nanoscience and Nanotechnology, The Hebrew University of Jerusalem, 91904 Jerusalem, Israel}

\author{Oded Millo}
\affiliation{Racah Institute of Physics and the Centre for Nanoscience and Nanotechnology, The Hebrew University of Jerusalem, 91904 Jerusalem, Israel}

\author{Ai-Min Guo}
\email[]{aimin.guo@csu.edu.cn}
\affiliation{Hunan Key Laboratory for Super-microstructure and Ultrafast Process, School of Physics and Electronics, Central South University, Changsha 410083, China}

\author{Qing-Feng Sun}
\affiliation{International Center for Quantum Materials, School of Physics, Peking University, Beijing 100871, China}
\affiliation{Collaborative Innovation Center of Quantum Matter, Beijing 100871, China}
\affiliation{CAS Center for Excellence in Topological Quantum Computation, University of Chinese Academy of Sciences, Beijing 100190, China}

\begin{abstract}

Recently, electron transport along chiral molecules has been attracting extensive interest and a number of intriguing phenomena have been reported in recent experiments, such as the emergence of zero-bias conductance peaks in the transmission spectrum upon the adsorption of single-helical protein on superconducting films. Here, we present a theoretical study of electron transport through a two-terminal single-helical protein sandwiched between a superconducting electrode and a normal-metal one in the presence of a perpendicular magnetic field. As the proximity-induced superconductivity attenuates with the distance from superconducting media, the pairing potential along the helix axis of the single-helical protein is expected to decrease exponentially, which is characterized by the decay exponent $\lambda$ and closely related to the experiments. Our results indicate that (i) a zero-bias conductance peak of $2e^2/h$ appears at zero temperature and the peak height (width) decreases (broadens) with increasing temperature, and (ii) this zero-bias peak can split into two peaks, which are in agreement with the experiments [see, e.g., Nano Lett. {\bf 19}, 5167 (2019)]. Remarkably, Majorana zero modes are observed in this protein-superconductor setup in a wide range of model parameters, as manifested by the $Z_2$ topological invariant and the Majoroana oscillation. Interestingly, a specific region is demonstrated for decaying superconductivity, where topologically nontrivial and trivial zero modes coexist and the bandgap remains constant. With increasing the pairing potential, the topologically nontrivial zero modes will transform to the trivial ones without any bandgap closing-reopening, and the critical pairing potential of the phase transition attenuates exponentially with $\lambda$. Additionally, one of the two zero modes can be continuously shifted from one end of the protein toward the other end contacted by the normal-metal electrode. The underlying physics of the topologically nontrivial and trivial zero modes is discussed.

\end{abstract}

\maketitle
\date{\today}

\section{\label{sec1}Introduction}

Majorana fermion, firstly predicted to be one fundamental particle by Ettore Majorana in \textrm{1937} \cite{Majorana}, was considered as an excitation in condensed matter physics. Unpaired Majorana fermions were reported on vortices of chiral two-dimensional $p$-wave superconductors \cite{N,Jackiw,Ivanov,L}. Kitaev put forward a seminal model to detect such exotic particles at the two ends of a nanowire, which may facilitate realizing half-qubits in topological quantum computing schemes \cite{Kitaev}. Since then, a number of one-dimensional systems were proposed to observe Majorana fermions or Majorana zero modes, such as ultracold fermionic atoms with spin-orbit interaction \cite{Jiang}, quantum dot chains \cite{Fulga,Nadj}, helical Shiba chains \cite{Yazdani,K}, and ferromagnetic atomic chains on superconducting substrates \cite{S,J,Ruby,Jeon,Kim}. Importantly, a possible signature of Majorana fermions is the emergence of zero-bias conductance peaks in the transmission spectrum, which were reported in diverse systems during the past decade \cite{Sau,Mourik,Zhang,Chiu}.

On the other hand, chiral molecules including double-stranded DNA (dsDNA) and single-helical protein have been receiving extensive attention. For instance, the spin selectivity effect was widely demonstrated in a variety of chiral molecules \cite{gb,Guo,Mishra,dgf,uy,qq}. Topological states were reported in both dsDNA and single-helical molecules under a perpendicular electric field, and a Thouless quantum pump could be realized by rotating this electric field in the transverse plane \cite{gam1,gam2}.

Recently, the chiral molecule-superconductor hybrid systems attracted intensive interest. As early as 2001, Kasumov \emph{et al.} firstly measured the electron transport through $\lambda$-DNA connected to superconducting electrodes in a magnetic field, demonstrating a zero-bias conductance peak and proximity-induced superconductivity \cite{Kasumov}. Then in 2016, Alpern \emph{et al.} fabricated a hybrid system where single-helical protein of $\alpha$-helical polyalanine self-assembled on a superconducting Nb film and observed a zero-bias conductance peak by using scanning tunneling spectroscopy \cite{Alpern}. They found that the peak height decreases with increasing the temperature and its width increases simultaneously, which may be related to unconventional triplet-pairing components with either $d$- or $p$-wave symmetry \cite{Alpern}. Recently, Millo \emph{et al.} studied the electron transport properties of several chiral molecule-superconductor systems by adsorbing $\alpha$-helical polyalanine on different conventional superconductors \cite{Shapira,Yavilberg}. They observed a robust zero-bias conductance peak in the transmission spectrum, finding that the peak height is reduced by increasing the temperature and this peak vanishes in the strong magnetic field regime, which points toward the topological triplet $p$-wave superconductivity \cite{Shapira,Yavilberg}. All these experiments share a common feature of zero-bias conductance peaks in the chiral molecule-superconductor systems and the underlying physics remains elusive. Theoretically, Tang \emph{et al.} investigated the topological properties of a single-stranded DNA deposited on an $s$-wave superconductor, finding Majorana zero modes in this hybrid system \cite{Tang}. Very recently, Chen \emph{et al.} studied the transport properties of a dsDNA proximity-coupled by an $s$-wave superconductor, demonstrating Majorana zero modes as well as topological phase transitions in this dsDNA-superconductor system \cite{Qiao}. Notice that all these theoretical works focus on DNA molecules contacted by two normal-metal electrodes under homogeneous superconductivity, which are completely different from previous experiments \cite{Kasumov,Alpern,Shapira,Yavilberg}.

In this paper, we study theoretically the topological properties of a single-helical protein, connected to a superconducting electrode and a normal-metal one, in the presence of a Zeeman field pointing toward the y axis, as illustrated in Fig.~\ref{fig:one}. In this protein-superconductor system, the pairing potential attenuates exponentially along the helix axis (z axis), which captures the main features of previous experiments \cite{Alpern,Shapira,Yavilberg} and is termed as decaying superconductivity here. We find that a zero-bias conductance peak of $2e^2/h$ is demonstrated at zero temperature and the peak height decreases with increasing the temperature which is accompanied by the peak width broadening, and this zero-bias peak splits into two peaks by increasing the Zeeman field, consistent with the experiments \cite{Alpern,Shapira,Yavilberg}. Besides the existence of Majorana zero modes in a wide range of model parameters, a specific region is observed in the case of decaying superconductivity, in which topologically nontrivial and trivial zero modes coexist and the bandgap remains unchanged. In particular, the topologically nontrivial zero modes can transform to the trivial ones in the absence of bandgap closing-reopening within this specific region, and the critical pairing potential of the phase transition decreases exponentially with the superconducting decay exponent $\lambda$. Instead of localized at the ends, one of the two zero modes could be localized at any position of the single-helical protein and be shifted from the left end toward the right one by increasing the pairing potential.

The rest of the paper is organized as follows. In Sec.~\ref{sec2}, we introduce the Hamiltonian of the single-helical protein and the Bogoliubov-de-Gennes (BdG) Hamiltonian and detail the methods for  our calculation. In Sec.~\ref{sec3},  the numerical results are displayed and the discussion is made. In Sec.~\ref{sec4}, a summary is given.

\section{\label{sec2}Model and Method}

\begin{figure}
\includegraphics[width=\columnwidth]{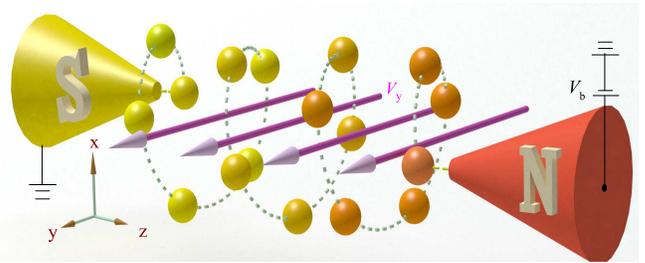}
\caption{\label{fig:one} Schematic diagram of a two-terminal single-helical protein, contacted by the left superconducting electrode S and the right normal-metal one N, in the presence of a perpendicular Zeeman field $V_{\rm y}$ which points along the ${\rm y}$ axis. Here, each sphere represents an amino acid and the dashed line connecting these spheres depicts the helical structure of the protein. Because of the proximity-induced superconductivity by the superconducting electrode, the pairing potential emerges simultaneously in the protein and decays exponentially along the helix axis, as illustrated by the gradient color.}
\end{figure}

We consider a single-helical protein connected to a superconducting electrode and a normal-metal one under a perpendicular magnetic field, as sketched in Fig.~\ref{fig:one}, which is similar to previous experiments \cite{Alpern,Shapira,Yavilberg}. The total Hamiltonian is written as
\begin{equation}
\mathcal{H}= \mathcal{H}_ {\text{BdG}}+ \mathcal{H} _{\text{L}}+ \mathcal{H} _{\text{R}} +\mathcal{H} _{\text{c}}. \label{eq1}
\end{equation}
Here, $\mathcal{H}_ {\text{BdG}}$ is the Hamiltonian of the single-helical protein with superconductivity, $\mathcal{H}_ {\text{L}}$ and $\mathcal{H}_ {\text{R}}$ are the Hamiltonians of the left superconducting electrode and the right normal-metal one, respectively, and $\mathcal{H}_ {\text{c}}$ describes the coupling between the electrodes and the protein. $\mathcal{H}_ {\text {BdG}}$ is expressed as
\begin{equation}
\mathcal{H}_ {\text{BdG}}= \frac 1 2\sum_{n,n'} \Psi_{n}^{\dag}\textbf{H}_{\text{BdG}}(n,n')\Psi_{n'}, \label{eq2}
\end{equation}
with the matrix element
\begin{equation}
\textbf{H}_{\text{BdG}}(n,n') =\left[ \begin{array}{cc}
\mathcal{H}_{\text{p}}(n,n') +\mu\delta_{n,n'} & -i\Delta_{n}{ \hat \sigma_ {\textrm{y}}}\delta_{n,n'} \\
i\Delta_{n}^ {\ast}{\hat \sigma_ {\textrm{y}}}\delta_{n,n'} & -\mathcal{H}_ {\text{p}}^{\ast}(n,n') -\mu\delta_{n,n'} \\
\end{array} \right]. \label{eq3}
\end{equation}
Here, $\Psi_n ^\dagger =(c_{n \uparrow} ^\dagger, c_{n \downarrow} ^\dagger, c_{n \uparrow}, c_{n \downarrow})$, $\mu$ is the chemical potential, and $\delta _{n, n'}$ the Kronecker delta. Because of the proximity-induced superconductivity by the left superconducting electrode, a pairing potential $\Delta_n$ emerges in the single-helical protein. Notice that since the superconducting wave-function attenuates exponentially with the distance from the superconducting electrode, this pairing potential along the helix axis may be expressed as $\Delta_{n}=\Delta e^{-n/\lambda}$ \cite{Sonin,Samoilenka,Volkov}. Here, $\Delta$ is the pairing potential in the superconducting electrode and $\lambda$ is the decay exponent which describes how fast the superconductivity attenuates along the protein. $\mathcal{H}_{\text{p}}$ is the Hamiltonian of the protein without any superconductivity and reads \cite{A}
\begin{equation}
\begin{split}
\mathcal{H}_ {\text{p}}= \sum_{n=1}^N \epsilon_n c_n^ {\dagger}
c_n + \sum_{n=1} ^{N-1} \sum_{m=1} ^{N-n} t_m c_n^ {\dagger} c_{n+m}
+\sum_ {n=1} ^N V_{\textrm{y}} c_n ^{\dagger} \hat {\sigma}_{\textrm{y}} c_n \\
+\sum_{n=1} ^{N-1} \sum_{m=1} ^{N-n}2 i s_m \cos (\psi_{n ,m}^-)c_n ^{\dagger} \hat {\sigma}_{n,m} c_{n+m}  +\text{H.c.}, \label{eq4}
\end{split}
\end{equation}
where $c_n ^{\dagger}=(c_{n \uparrow} ^{\dagger}, c_ {n\downarrow} ^{\dagger})$ is the creation operator at site $n$ of the protein with length $N$. $\epsilon_n$ is the on-site energy, $t_m= t_1 e^{-( l_m- l_1)/ l_c}$ the $m$th neighboring hopping integral, $V_{\textrm{y}}$ the Zeeman field strength, and $s_m=s_1 e^{-(l_m -l_1)/l_c}$ the renormalized spin-orbit coupling \cite{A}. Here, $l_m= \sqrt{ [2R \sin(m \Delta \psi/2)] ^{2}+ (m\Delta h)^2}$ is the Euclidean distance between sites $n$ and $n+m$, $l_c$ is the other decay exponent which characterizes the long-range electron transport in the protein, $R$ the radius, and $\Delta \psi$ $(\Delta h)$ the twist angle (stacking distance) between the nearest neighboring sites. $\hat {\sigma} _{n,m}= (\hat {\sigma}_ {\textrm{x}}\sin \psi_{n,m}^{+}- \hat{\sigma}_ {\textrm{y}} \cos \psi_{n,m} ^{+})\sin \theta_{m} +\hat{\sigma} _{\textrm{z}} \cos\theta_{m}$, $\theta_{m} =\arccos[2R \sin(m\Delta \psi/2) /l_m]$, $\psi_{n,m}^ {\pm}=(\psi_{n+m} \pm \psi_{n})/2$, $\psi_n=n \Delta \psi$, and $\hat{\sigma}_{{\rm x,y,z}}$ the pauli matrices.

The Hamiltonians of the two electrodes and their coupling to the protein is written as
\begin{equation}
\mathcal{H}_{\text{L}}=\sum_{k} \epsilon_ {k\text{L}}a_{k\text{L}} ^{\dagger} a_{k\text{L}}+\sum_{k}(\Delta a_{k\uparrow \text{L}} ^{\dagger} a_{-k\downarrow \text{L}} ^{\dagger}+\Delta a_{-k\downarrow \text{L}}a_{k\uparrow \text{L}}),  \label{eq5}
\end{equation}
\begin{equation}
\mathcal{H}_{\text{R}} =\sum_{k}\epsilon_ {k\text{R}} a_{k\text{R}} ^{\dagger} a_{k\text{R}}, \label{eq6}
\end{equation}
and
\begin{equation}
\mathcal{H}_ {\text{c}}= \sum_{k} (t a_{k \text{L}} ^{\dagger} c_{1}+t a_{k \text{R}}^{\dagger}c_{N}+\text{H.c.}). \label{eq7}
\end{equation}
Here, $a_{k \alpha} ^{\dagger}=(a_{k\uparrow \alpha} ^{\dagger}, a_{k \downarrow \alpha} ^{\dagger})$ is the creation operator for an electron with wave vector $k$ in the left and right electrodes ($\alpha=\text{L}, \text{R} $), and $t$ is the coupling between the protein and the electrodes. We emphasize that the left electrode is superconducting with the pairing potential $\Delta$ and the right one is normal metal. The retarded Green's function can be expressed as
\begin{equation}
{\textbf{G}^{r}} (E)=[E \textbf{\textbf{I}}- \textbf{H} _{\text{BdG}}- {\bf \Sigma}_ {\text{R}} ^{r}(E)-{\bf \Sigma}_ {\text{L}} ^{r}(E)]^{-1}, \label{eq8}
\end{equation}
where $E$ is the electron energy and $\textbf{I}$ the identity matrix of size $4N \times 4N$. For the normal-metal electrode, the retarded self-energy is ${\Sigma}_ {\text{R}} ^{r}(E)=-i {\Gamma}_{0}/{2}$ with ${\Gamma}_{0} =2\pi t^{2} \Sigma_{k} \delta(E-\epsilon_{k \alpha})$. While for the superconducting electrode, the retarded self-energy reads \cite{addr1,addr2,Yang,Pan}
\begin{equation}
 {\bf\Sigma}_{\text{L}}^{r}(E)=-i{\Gamma}_{0}\beta(E)\left( \begin{array}{cc}
 1 & \Delta/E\\
 \Delta/E & 1 \\
\end{array} \right), \label{eq9}
\end{equation}
where $\beta(E)= |E|/\sqrt{E^2 -\Delta^2}$ for $|E|> \Delta$ and $\beta(E) =E/i\sqrt {\Delta^2 -E^2}$ for $|E|<\Delta$. From the Keldysh formula, the lesser Green's function can be written as
\begin{equation}
{\bf G}^{<}(E)={\bf G} ^{r}(E)( {\bf\Sigma}_{\text{L}} ^{<}(E,V_ {\textrm{b} })+ {\bf\Sigma} _{\text{R}} ^{<}(E,V_ {\textrm{b}}) ){\bf G}^{a}(E). \label{eq10}
\end{equation}
Here, ${\bf\Sigma}_ {\alpha}^{<} (E,V_{\textrm{b}})$ is the lesser self-energy in the left ($\alpha=\text{L}$) and right ($\alpha= \text{R} $) electrodes, which can be evaluated from the retarded and advanced self-energies \cite{Tang}, with $V_{\rm b}$ the bias voltage. Then, the current flowing through the protein can be expressed as \cite{Sun,Cao}
\begin{equation}
\begin{split}
I_{\text{R}}=\frac e h\int \mathrm{d}E{\rm ReTr}\{\sigma[{\bf G}^{<}(E,V_{\textrm{b}}){\bf\Sigma}_{\text{R}}^{a}(E) \\
+{\bf G}^{r}(E){\bf\Sigma}_{\text{R}}^{<}(E,V_{\textrm{b}})]\}, \label{eq11}
\end{split}
\end{equation}
where $\sigma=diag(1,1,-1,-1\cdots)$ accounts for different charges carried by electrons and holes. Finally, the differential conductance can be evaluated as
\begin{equation}
G=\frac{d I_ {\text{R}}} {dV_{\textrm{b}}}. \label{eq12}
\end{equation}

To elucidate the topological properties of this chiral molecule-superconductor system, the $Z_2$ topological invariant is calculated by using the scattering matrix \cite{M, Akhmerov}
\begin{equation}
\textbf{S}=\left( \begin{array}{cc}
\textbf{R} & \textbf{T}^{'}\\
\textbf{T} & \textbf{R}^{'} \\
\end{array} \right)= \textbf{I}-2\pi i\textbf{W}^{\dagger}{\bf G}^{r}(E)\textbf{W}. \label{eq13}
\end{equation}
Here, $\textbf{W}$ can be obtained from $\mathcal{H}_{\text{c}}$, $\textbf{R}$ is the reflection matrix, and $\textbf{T}$ the transmission matrix. Subsequently, the topological number can be derived from the reflection matrix as $Q={\rm sign}[{\rm det}(\textbf{R})]$, and the $Z_{\textrm{2}}$  invariant $\nu$ satisfies $(-1)^{\nu}=Q$.

\section{\label{sec3}Results and Discussion}

For the single-helical protein, the on-site energy is set to $\epsilon_n =0$ without loss of generality, the nearest-neighbor hopping integral $t_1$ is taken as the energy unit, and the renormalized spin-orbit coupling as $s_1= 0.12t_1$ \cite{A}. The chemical potential is chosen as $\mu=1.7 t_1$ and the Zeeman field as $V_{\textrm{y}}= 1.5t_1$. The structural parameters are set to $R=0.25$ nm, $\Delta h=0.15$ nm, $\Delta \psi=5\pi/9$, and $N=80$. Then, the Euclidean distance between two nearest-neighboring sites is approximated as $l_1\sim 0.41$ nm, and the decay exponent characterizing the long-range electron transport in the protein is fixed as $l_c= 0.09$ nm \cite{A}. For the normal-metal electrode, the coupling strength is set to $\Gamma_0 =0.05t_1$. While for the superconducting electrode, the pairing potential $\Delta= 0.8t_1$. As the superconducting coherence length could range from tens of nanometers to hundreds of nanometers at low temperatures \cite{Katzir}, the decay exponent simulating how the superconductivity attenuates along the protein is set to $\lambda=100$. The values of all above-mentioned parameters will be used throughout the paper and all the calculations are performed at zero temperature, unless stated otherwise.

\begin{figure}
\includegraphics[width=0.8\columnwidth]{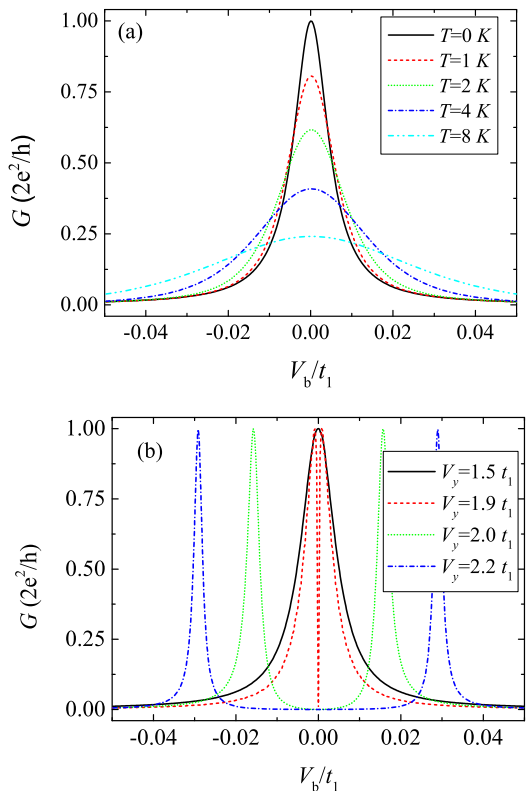}
\caption{\label{fig:two} Electron transport along a two-terminal single-helical protein by considering different temperatures $T$ and Zeeman fields $V_{\rm y}$. Differential conductance $G$ versus bias voltage $V_{\rm b}$ for (a) various $T$ with $V_{\rm y}= 1.5t_1$ and for (b) several $V_{\rm y}$ at zero temperature. Here, the model parameters are $N=80$, $\Delta=0.8 t_1$, $\mu= 1.7t_1$, and $\lambda= 100$. }
\end{figure}

We first consider how the temperature $T$ and the Zeeman field $V_{\rm y}$ affect the charge transport properties of the single-helical protein with decaying superconductivity, which will be termed as the protein-superconductor system in the following. Figure~\ref{fig:two}(a) shows the differential conductance $G$ versus the bias voltage $V_{\textrm{b}}$ by considering several values of $T$. A zero-bias conductance peak of $2e^2/h$ is clearly demonstrated in the transmission spectrum at zero temperature [see the black-solid line in Fig.~\ref{fig:two}(a)], indicating the existence of two zero modes in the protein-superconductor system. When the temperature is increased from $T=0$ K to $8$ K, the peak height is progressively declined from $G=2e^2/h$ to approximately $0.25e^2/h$, which is accompanied by the broadening of the peak width [see the other lines in Fig.~\ref{fig:two}(a)]. This conductance peak will finally vanish by further increasing $T$. These results are in good agreement with previous experiments that the peak height (width) decreases (broadens) with increasing $T$ \cite{Alpern,Yavilberg}.

\begin{figure*}
\includegraphics[width=2\columnwidth]{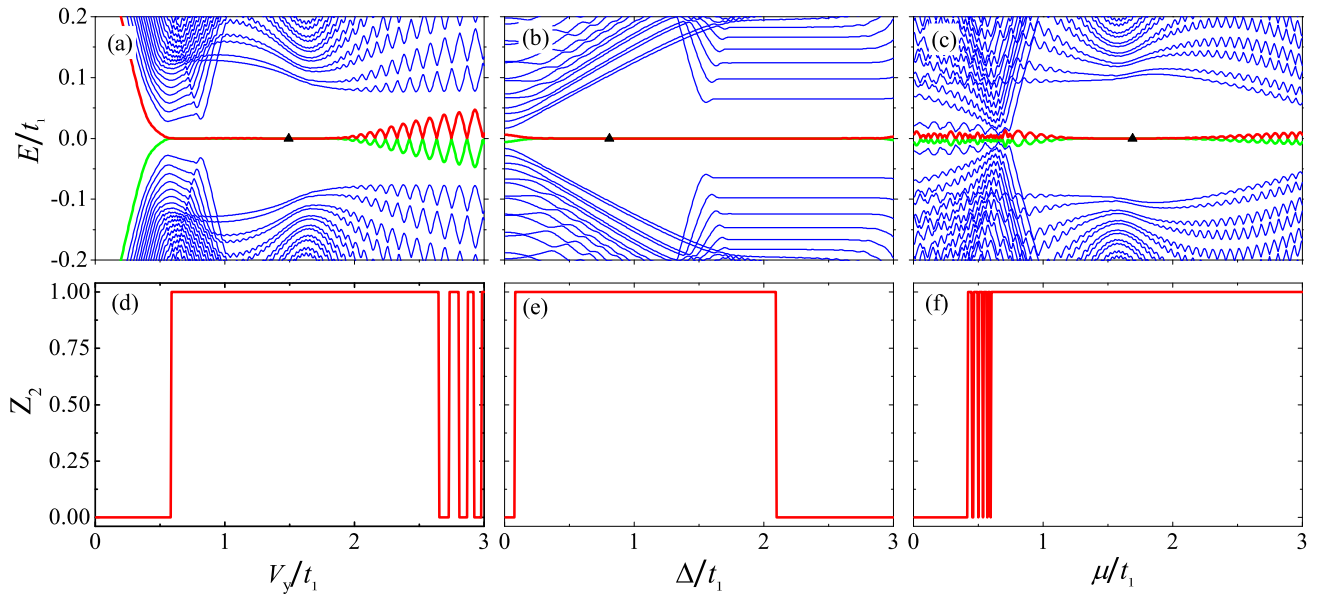}
\caption{\label{fig:three} Energy spectra and $Z_2$ topological invariants of the protein-superconductor system by changing the Zeeman field $V_{\rm y}$, the pairing potential $\Delta$, and the chemical potential $\mu$. (a)-(c) show the energy spectra as functions of $V_{\rm y}$, $\Delta$, and $\mu$, respectively. (d)-(f) display the $Z_2$ invariants as functions of $V_{\rm y}$, $\Delta$, and $\mu$, respectively. The triangles in (a)-(c) mark the point where the differential conductance in Fig.~\ref{fig:two}(a) is calculated.}
\end{figure*}

The above evolution phenomenon of the conductance peak with the temperature originates from the thermal broadening effect. We know that the electron distribution satisfies the Fermi-Dirac statistics, and all the electronic states of the electrode locating within the bias window $[-V_{\rm b}/2, V_{\rm b}/2]$ will contribute to the current $I_{\rm R}$ [see Eq.~(\ref{eq11})] and thus to $G$. At zero temperature, the Fermi-Dirac distribution function is a step function and only the electronic states at the Fermi level of the electrode can affect $G$. Whereas at nonzero temperatures, the Fermi-Dirac distribution function has hyperbolic-tangent-like form and consequently the contribution of the electronic states at the Fermi level will be decreased. Instead, both electronic states below and above the Fermi level will influence $G$, giving rise to the thermal broadening effect. As a result, the peak height decreases with increasing $T$ and consequently its width becomes wider. Since we consider an extremely small bias voltage between the left and right electrodes, only the electronic states of the protein-superconductor system around the Fermi level, i.e., the two zero modes [see the red- and green-solid lines in Fig.~\ref{fig:three}(a)], will attend the charge transport process. Therefore, the integral of $G$ along the $V_{\rm b}$ axis remains unchanged for different temperatures [Fig.~\ref{fig:two}(a)].

We then study the effect of the Zeeman field, as shown in Fig.~\ref{fig:two}(b) which plots the bias voltage-dependent $G$ at zero temperature for different $V_{\rm y}$. It clearly appears that the conductance peak locates at zero bias voltage for relatively small $V_{\rm y}$ [see the black-solid line in Fig.~\ref{fig:two}(b)]. In contrast, this zero-bias conductance peak will disappear for large $V_{\rm b}$ and instead a pair of peaks emerge symmetrically with respect to the line $V_{\rm b}=0$ [see the red-dashed, green-dotted, and blue-dash-dotted lines in Fig.~\ref{fig:two}(b)], which is attributed to the Majorana oscillation, as discussed below. These results are qualitatively consistent with the previous experiment that a zero-bias conductance peak appears for small magnetic field and splits into two peaks by increasing this magnetic field \cite{Yavilberg}. However, another experiment reported that the zero-bias conductance peak does not split with increasing the magnetic field upon adsorption of $\alpha$-helical polyalanine on Au films grown on superconducting NbN \cite{Shapira}, which may be related to strong hybridization effects at the chiral molecule-Au interface and deserve further investigations. Although the zero-bias conductance peak suggests the existence of zero modes in the protein-superconductor hybrid system, its physical origin remains unclear.

To understand the physical nature of the zero-bias conductance peak and the possible topological properties of the protein-superconductor system, the energy spectrum and the corresponding topological invariant are calculated by changing the Zeeman field $V_{\rm y}$, the pairing potential $\Delta$, and the chemical potential $\mu$. Figures~\ref{fig:three}(a) and~\ref{fig:three}(d) display the energy spectrum and the topological invariant, respectively, as a function of $V_{\rm y}$. One can see from Fig.~\ref{fig:three}(a) that the energy spectrum is always symmetric with respect to the line $E=0$, owing to the particle-hole symmetry. In relatively weak Zeeman field regime, a bandgap is clearly demonstrated due to the proximity-induced superconductivity and its width decreases almost linearly with increasing $V_{\rm y}$ [see the red- and green-solid lines in Fig.~\ref{fig:three}(a)]. This bandgap closes and reopens at $V_{\rm y}\sim 0.59 t_1$, implying a possible topological phase transition at this critical Zeeman field. This phase transition is further confirmed in Fig.~\ref{fig:three}(d) where the topological invariant changes from $Z_2 =0$ to $1$ at $V_{\rm y}\sim 0.59 t_1$. In the intermediate regime of $0.59t_1 <V_{\rm y}< 1.78t_1$, the two modes closest to zero energy overlap at $E=0$ [see the red- and green-solid lines in Fig.~\ref{fig:three}(a)] and two Majorana zero modes with $Z_2 =1$ appear in the energy spectrum [Fig.~\ref{fig:three}(d)], leading to the existence of the zero-bias conductance peak [Fig.~\ref{fig:two}(a)]. While in the large Zeeman field regime of $V_{\rm y}>1.78 t_1$, the two Majorana modes oscillate with increasing $V_{\rm y}$ and the oscillating amplitude increases almost linearly [see the right half part of Fig.~\ref{fig:three}(a)], which gives rise to the splitting of the zero-bias conductance peak into two peaks [Fig.~\ref{fig:two}(b)]. This phenomenon is named as the Majorana oscillation \cite{Rainis,Pan} and regarded as a smoking gun for the existence of Majorana modes \cite{Sarma}. Indeed, the Majorana oscillation of small amplitude corresponds to the topologically nontrivial modes with $Z_2 =1$. Interestingly, the oscillation of large amplitude refers to either the topologically nontrivial modes or the trivial ones, with the topological invariant oscillating between $Z_2 =0$ and $1$ for $V_{\rm y}> 2.65t_1$ [see the rightmost part of Fig.~\ref{fig:three}(d)], indicating the emergence of multiple topological phase transitions in the large Zeeman field regime.

We then investigate the effect of the superconducting pairing potential $\Delta$ on the energy spectrum and the topological invariant, as presented in Figs.~\ref{fig:three}(b) and~\ref{fig:three}(e), respectively. It is clear that there does not exist a bandgap in the absence of pairing potential, and the two modes closest to $E=0$ seem to be separated from each other at $\Delta=0$ [see the red- and green-solid lines in Fig.~\ref{fig:three}(b)], owing to finite-size effects. When the pairing potential is incorporated, a bandgap appears and its width presents nonmonotonic dependence on $\Delta$. This gap width firstly increases with $\Delta$ and then decreases with $\Delta$ [see the left part of Fig.~\ref{fig:three}(b)]. Interestingly, this bandgap remains unchanged for $1.60 t_1 < \Delta< 2.95 t_1$ [see the right half part of Fig.~\ref{fig:three}(b)], of which this range is named as specific region. We stress that several important phenomena can be identified in this specific region, as further discussed below. In addition, the two modes quickly overlap at $E=0$ with increasing $\Delta$ and two zero modes appear in a very wide range of $\Delta$, which can be divided into two types as characterized by the topological invariant. The topological invariant changes from $Z_2 =0$ for $\Delta <0.08t_1$ to $Z_2 =1$ for $0.08t_1 <\Delta <2.09t_1$, and back to $Z_2 =0$ for $\Delta >2.09t_1$ [Fig.~\ref{fig:three}(e)], implying a topological phase transition at the critical values of $\Delta \sim 0.08t_1$ and $2.09t_1$. It is interesting that beyond the Majorana zero modes for $\Delta <2.09t_1$, topologically trivial zero modes also exist in the protein-superconductor system for $\Delta >2.09t_1$. We point out that these topologically trivial zero modes can generate zero-bias conductance peaks in the transmission spectrum as well, just as the Majorana zero modes.

\begin{figure*}
\includegraphics[width=2\columnwidth]{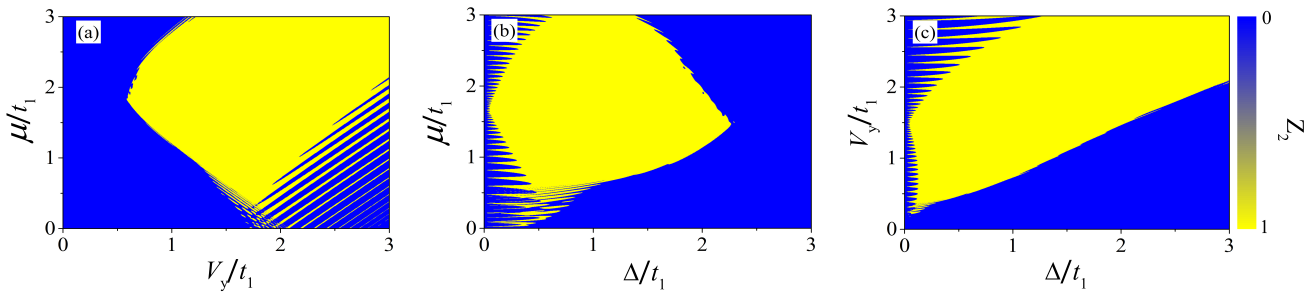}
\caption{\label{fig:four} Phase diagrams of the protein-superconductor system in the parameter space of the Zeeman field $V_{\rm y}$, the pairing potential $\Delta$, and the chemical potential $\mu$. (a) $Z_2$ topological invariant versus $V_{\rm y}$ and $\mu$. (b) $Z_2$ versus $\Delta$ and $\mu$. (c) $Z_2$ versus $\Delta$ and $V_{\rm y}$. The remaining parameters are the same as those in Fig.~\ref{fig:two}.}
\end{figure*}

Figures~\ref{fig:three}(c) and~\ref{fig:three}(f) show the energy spectrum and the topological invariant, respectively, as a function of $\mu$. One clearly identifies the oscillation phenomenon that the two modes closest to $E=0$ oscillate sharply with increasing $\mu$ for $\mu <1.34t_1$, and the oscillation pattern is significantly modified across the critical value of $\mu \sim0.74t_1$ at which a bandgap takes place [see the red- and green-solid lines in Fig.~\ref{fig:three}(c)]. In the low chemical potential regime of $\mu< 0.42t_1$, although the oscillating amplitude is small, the topological invariant is $Z_2 =0$ [Fig.~\ref{fig:three}(f)]and consequently the two modes are topologically trivial. For $0.42 t_1 < \mu< 0.60t_1$, the topological invariant oscillates dramatically between $Z_2=0$ and $1$, in which both topologically nontrivial and trivial modes exist. While for $ 0.60t_1 <\mu< 1.34t_1$, the topological invariant becomes $Z_2 =1$ and the two modes are topologically nontrivial. By further increasing $\mu$, the two modes become degenerated at $E=0$ with $Z_2 =1$ and the Majorana zero modes emerge for $1.34t_1 <\mu <1.91t_1$. In the high chemical potential regime of $\mu> 1.91t_1$, the two modes oscillate around $E=0$ again and the Majorana oscillation is identified with $Z_2 =1$, where the topologically nontrivial zero and nonzero modes coexist.

We then investigate the topological phase diagrams of the protein-superconductor system in different parameter spaces by taking into account $V_{\rm y}$, $\Delta$, and $\mu$. Figure~\ref{fig:four}(a) shows the topological invariant as functions of $V_{\rm y}$ and $\mu$. One can see that the topological invariant is $Z_2 =0$ for relatively weak Zeeman fields [see the leftmost blue part in Fig.~\ref{fig:four}(a)] and thus all the modes are topologically trivial, which is independent of $\mu$. When the Zeeman field is increased up to $V_{\rm y} \sim 0.58t_1$, the topological invariant could transform to $Z_2 =1$ and gives rise to the Majorana modes. This is consistent with the previous work that a sufficiently large Zeeman field is necessary to drive effective $p$-wave superconducting states into topological superconducting phase \cite{Kane}, where the $p$-wave superconductivity in the single-helical protein could be induced by the combination of spin-orbit coupling and pairing potential \cite{R}. With increasing $V_{\rm y}$, the range of $\mu$ supporting the Majorana modes with $Z_2 =1$ is enlarged almost linearly [see the yellow part in Fig.~\ref{fig:four}(a)], which is accompanied by the topological phase transition at the blue-yellow border. Remarkably, a number of yellow strips are found in the $V_{\rm y}-\mu$ space, especially for relatively strong Zeeman fields and low chemical potentials [see the bottom-right part in Fig.~\ref{fig:four}(a)], because of the Majorana oscillation. It is clear that the topological invariant oscillates between $Z_2 =0$ and $1$ by increasing either $V_{\rm y}$ or $\mu$, implying multiple topological phase transitions.

Figures~\ref{fig:four}(b) and~\ref{fig:four}(c) plot the topological invariant in the $\Delta-\mu$ space and the $\Delta-V_{\rm y}$ space, respectively. It can be seen that all the modes are topologically trivial for extremely small $\Delta$, irrespective of $\mu$ and $V_{\rm y}$. In other words, a finite pairing potential is necessary to ensure topologically nontrivial phases. When the pairing potential reaches the critical value of $\Delta \sim 0.02t_1$, a number of discrete Majorana modes are identified in both the $\Delta-\mu$ space and the $\Delta-V_{\rm y}$ space. And the Majorana oscillation emerges simultaneously, where the topological invariant oscillates with increasing $\mu$ or $V_{\rm y}$ albeit of no oscillation phenomenon by changing $\Delta$. In the $\Delta-\mu$ space, the chemical potential range of the Majorana modes decreases with $\Delta$ and all the modes return to the topologically trivial ones when $\Delta >2.30t_1$, which is accompanied by the disappearance of the Majorana oscillation in the large pairing potential regime, as shown in Fig.~\ref{fig:four}(b). Contrarily, both the Majorana oscillation and the Majorana modes remain for large $\Delta$ in the $\Delta-V_{\rm y}$ space [Fig.~\ref{fig:four}(c)]. Additionally, the Majorana modes could exist at extremely low even zero chemical potential [Figs.~\ref{fig:four}(a) and~\ref{fig:four}(b)] and at weak Zeeman fields of $V_{\rm y}\sim0.18t_1$ [Fig.~\ref{fig:four}(c)]. Therefore, we conclude that the Majorana modes can be observed in the protein-superconductor system in a very wide range of the Zeeman field, the pairing potential, and the chemical potential. We stress that the phase diagrams of the protein-superconductor system are different from previous works on single-stranded DNA \cite{Tang}, dsDNA \cite{Qiao}, multichannel Majorana nanowires \cite{Alvarado}, and antiferromagnetic chains \cite{Sedlmayr}, which mainly originates from the decaying superconductivity.

We have demonstrated above the topologically nontrivial zero modes as well as the trivial ones in the single-helical protein contacted by the left superconducting electrode and the right normal-metal electrode. Nevertheless, the physical origin of these trivial zero modes and the influence of the decaying superconductivity remain unclear. In the following, we investigate the topological properties of the protein-superconductor system by taking into account several decay exponents $\lambda$.

\begin{figure*}
\includegraphics[width=2\columnwidth]{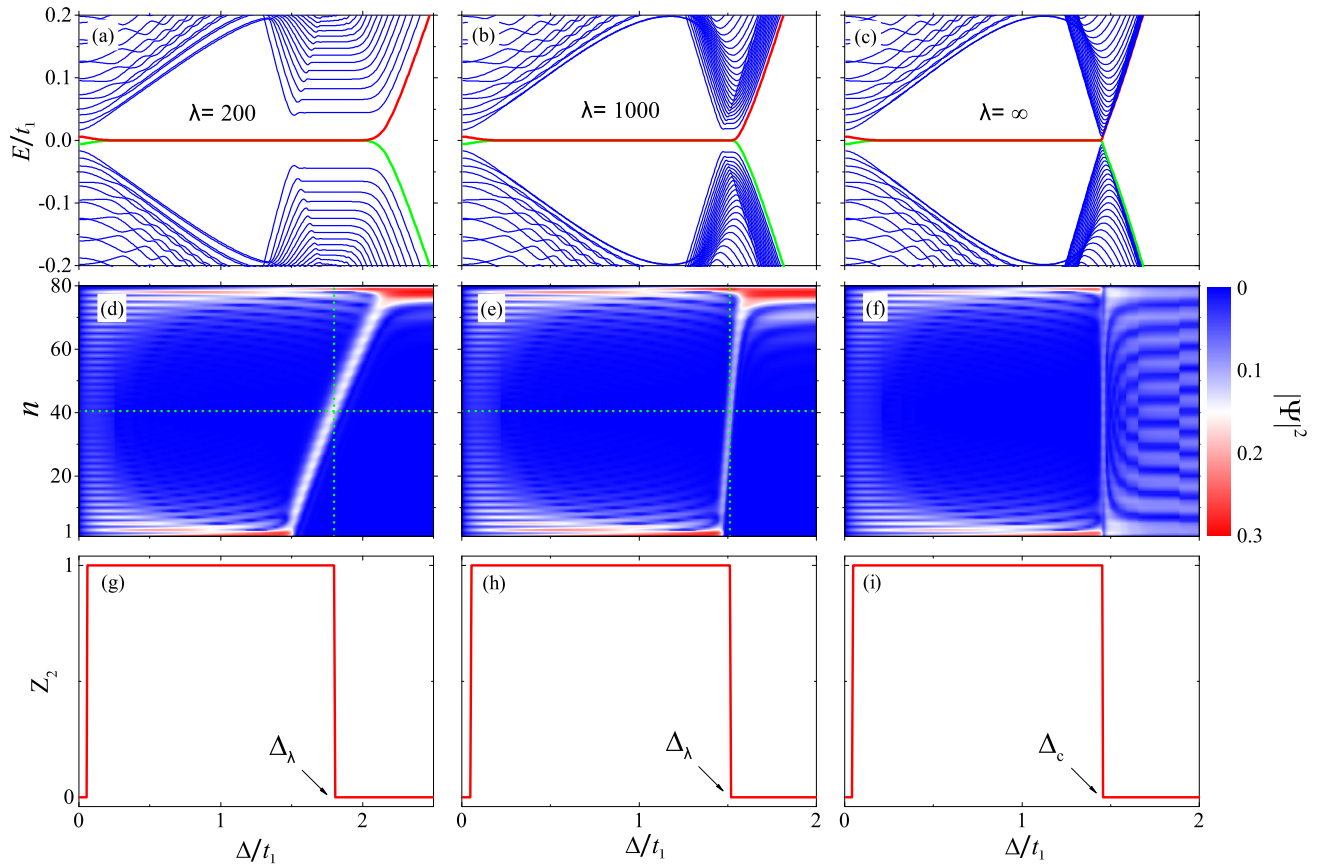}
\caption{\label{fig:five} Energy spectra, wave-functions, and $Z_2$ topological invariants of the protein-superconductor system by considering both decaying and homogeneous superconductivity, as a function of the pairing potential $\Delta$. (a)-(c) show the energy spectra versus $\Delta$ for $\lambda=200$, $1000$, and $\infty$, respectively. (d)-(f) show the spatial distribution of wave-functions $|\Psi| ^2$ of the zero modes for $\lambda=200$, $1000$, and $\infty$, respectively, as a function of $\Delta$. (g)-(i) display the corresponding $Z_2$ invariants versus $\Delta$. Here, the vertical and horizontal cyan-dotted lines in (d) and (e) denote the lines $\Delta =\Delta_ \lambda$ and $n= (N+1)/2$, respectively. $\Delta_ \lambda$ corresponds to the topological phase transition point in the case of decaying superconductivity, and $\Delta_{\rm c}$ to the homogeneous one. The other parameters are the same as those in Fig.~\ref{fig:two}.}
\end{figure*}

Figures~\ref{fig:five}(a) and~\ref{fig:five}(b) show the energy spectrum versus $\Delta$ with the decay exponent $\lambda=200$ and $1000$, respectively. As a comparison, Fig.~\ref{fig:five}(c) displays the energy spectrum by considering homogeneous superconductivity along the single-helical protein, i.e., $\lambda=\infty$. The energy spectra are similar for different values of $\lambda$, including the evolution of the bandgap and the two modes closest to zero energy, as shown in Figs.~\ref{fig:three}(b) and~\ref{fig:five}(a)-\ref{fig:five}(c). Interestingly, there always exists a specific region in the protein-superconductor system with the decaying superconductivity, where the gap width remains constant by changing $\Delta$. The range of this specific region shrinks from $[1.60t_1, 2.95t_1]$ for $\lambda=100$ to $[1.55t_1, 2.05t_1]$ for $\lambda=200$, and to $[1.48t_1, 1.54t_1]$ for $\lambda =1000$, and the associated bandgap decreases with increasing $\lambda$. In contrast, this specific region vanishes in the case of homogeneous superconductivity and is reduced to a single point at which the bandgap closes and reopens [Fig.~\ref{fig:five}(c)]. Since the bandgap closing-reopening process refers to a topological phase transition \cite{X}, intriguing phenomena may be expected in this specific region, as shown below. Besides, the two zero modes will be separated from each other at large pairing potential and finally the bandgap increases with $\Delta$.

Figures~\ref{fig:five}(d)-\ref{fig:five}(f) show the spatial distribution of wave-function $|\Psi|^2$ of the zero modes versus $\Delta$ with $\lambda =200$, $1000$, and $\infty$, respectively, while Figs.~\ref{fig:five}(g)-\ref{fig:five}(i) plot the corresponding topological invariants. At extremely small pairing potential regime, the wave-function could be delocalized over the whole system for both decaying and homogeneous superconductivity [see the leftmost parts of Figs.~\ref{fig:five}(d)-\ref{fig:five}(f)], and the associated topological invariant is $Z_2 =0$, owing to finite-size effects and weak proximity-induced superconductivity. We point out that the wave-function will be localized at the two ends of the protein for long molecular length. With increasing $\Delta$, the Majorana zero modes with $Z_2 =1$ appear and the wave-function is localized at the two ends as expected, for whatever the values of $\lambda$. It is surprising that in the case of decaying superconductivity, the localized wave-function can gradually move from the left end contacted by the superconducting electrode to the right end with increasing $\Delta$ when the pairing potential surpasses the critical value [see the white-oblique trajectories in Figs.~\ref{fig:five}(d) and~\ref{fig:five}(e)]. In other words, the wave-function can be localized at any position of the protein by properly tuning $\Delta$. Here, the moving of the localized wave-function is termed as the migration process, which only occurs in the aforementioned specific region of the protein-superconductor system with the decaying superconductivity. This migration process would be beneficial for implementing the braiding of the Majorana zero modes and for constructing the topological qubits \cite{addr3,addr4}.


Besides the tunable position of the localized wave function, we then discuss the other interesting phenomena in this specific region. It is well-known that the topological phase transition is usually accompanied by the bandgap closing-reopening \cite{X}, as further confirmed in Figs.~\ref{fig:five}(c) and~\ref{fig:five}(i) where the bandgap closes and reopens at $\Delta_{\rm c} \sim 1.45t_1$ with the topological invariant changing from $Z_2 =1$ to $0$. However, this picture is completely modified by considering the decaying superconductivity. During the migration process in which the bandgap remains constant, the topological invariant can transform from $Z_2 =1$ to $0$ at the critical value $\Delta_\lambda$, indicating a topological phase transition at $\Delta_\lambda$. This critical pairing potential decreases from $\Delta_ \lambda \sim 2.09t_1$ for $\lambda =100$ to $\Delta_ \lambda \sim 1.80t_1$ for $\lambda =200$, and to $\Delta_ \lambda \sim 1.51t_1$ for $\lambda =1000$. Interestingly, the dependence of $\Delta_\lambda$ on $\lambda$ can be approximated as
\begin{equation}
\Delta_ \lambda \sim \Delta_{\rm c} e^{(N+1) /2\lambda}. \label{eq14}
\end{equation}
We find that the crossing point between the lines $\Delta= \Delta_\lambda$ and $n=(N+1)/2$ locates in the white-oblique trajectories [see the cyan-dotted lines in Figs.~\ref{fig:five}(d) and~\ref{fig:five}(e)], which may provide an experimentally accessible way to discern the topologically nontrivial and trivial zero modes. For the Majorana zero modes, the wave-function is localized at the left half part and the right end of the protein; whereas for the topologically trivial zero modes, it is localized at the right half part and the right end. By further increasing $\Delta$, the wave-function is localized at the right end for the decaying superconductivity [see the rightmost parts of Figs.~\ref{fig:five}(d) and~\ref{fig:five}(e)], in sharp contrast to the homogenous case that the wave-function will be delocalized over the whole system [see the rightmost part of Fig.~\ref{fig:five}(f)].

Although we have demonstrated the topologically nontrivial and trivial zero modes in the single-helical protein when the superconductivity decays exponentially along the helix axis, two issues remain unclear: (i) the relationship between the migration process [see the white-oblique trajectories in Figs.~\ref{fig:five}(d) and~\ref{fig:five}(e)] and the decaying superconductivity, and (ii) why the phase transition point decreases with increasing the decay exponent $\lambda$? We will attempt to solve these issues in the following.

\begin{figure}
\includegraphics[width=\columnwidth]{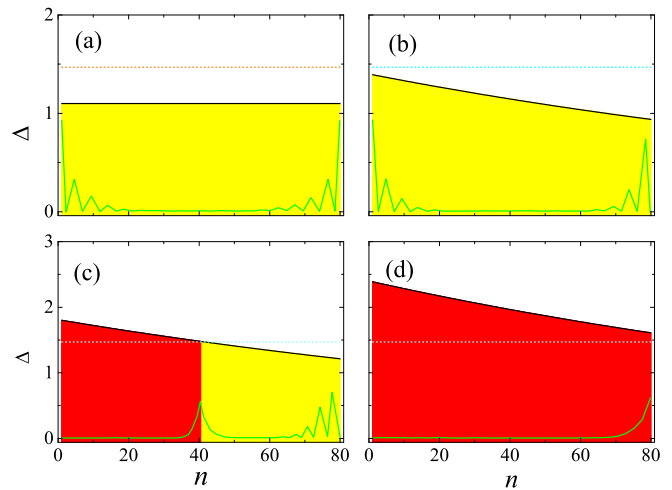}
\caption{\label{fig:six} Schematics of the topological properties and the spatial distribution of wave-function $|\Psi| ^2$ in the protein-superconductor system by considering both decaying and homogeneous superconductivity. (a) Homogeneous superconductivity with all the pairing potentials $\Delta_n=\Delta$ and $\Delta <\Delta_{\rm c}$. Decaying superconductivity with the pairing potential satisfying (b) $\Delta_1 <\Delta_ {\rm c}$, (c) $\Delta_N <\Delta_{\rm c} <\Delta_1$, and (d) $\Delta_N >\Delta_{\rm c}$. Notice that the pairing potential along the helix axis attenuates exponentially and thus $\Delta _N <\Delta_{N-1} <\cdots <\Delta_1$. Here, the black-solid lines describe $\Delta_n$ at site $n$ of the single-helical protein and the green-solid ones show $|\Psi| ^2$. The orange- and cyan-dotted lines denote the critical pairing potential $\Delta_{\rm c}$ for the homogeneous superconductivity, at which the topological phase transition takes place. The yellow regions refer to the topologically nontrivial phase and the red ones to the trivial phase.}
\end{figure}

Figure~\ref{fig:six}(a)-\ref{fig:six}(d) show the schematic diagrams of the relation between the topological phase and the pairing potential in the protein-superconductor system, where the first panel corresponds to the homogeneous superconductivity and the remaining ones to the decaying superconductivity. Here, the black-solid lines describe the evolution of the pairing potential along the single-helical protein and the green-solid ones show the wave function $|\Psi| ^2$. All the orange- and cyan-dotted lines denote the critical pairing potential $\Delta_{\rm c}$ for the homogeneous  superconductivity, where the topological phase transition occurs. The yellow regions refer to the topologically nontrivial phase and the red ones to the trivial phase. For the homogeneous superconductivity, i.e., $\Delta_n =\Delta$, the Majorana zero modes appear when $\Delta_n < \Delta_{\rm c}$, as characterized by the yellow region in Fig.~\ref{fig:six}(a). Correspondingly, the zero modes are localized at both ends of the protein and the wave-function is symmetric with respect to the line $n=(N+1)/2$ [see the black-solid line in Fig.~\ref{fig:six}(a)]. When $\Delta_n > \Delta_{\rm c}$, a topological phase transition takes place and all the modes becomes topologically trivial, which will be characterized the red region (not shown here).

We then consider the pairing potential decays exponentially along the protein, i.e., $\Delta_n =\Delta e^{-n/\lambda}$. In the limit case $\Delta_1< \Delta_{\rm c}$ where all the pairing potentials in the protein are smaller than $\Delta_{\rm c}$, the Majorana zero modes will emerge [see the yellow region in Fig.~\ref{fig:six}(b)] and be localized at both ends as well albeit of asymmetric wave function [see the green-solid line in Fig.~\ref{fig:six}(b)]. In the other limit case $\Delta_N> \Delta_{\rm c}$ that all the pairing potentials are larger than $\Delta_{\rm c}$, the Majorana zero modes transform to the topologically trivial ones and will be localized at the right end [see the red region and the green-solid line in Fig.~\ref{fig:six}(d)].

Of particular interest is the intermediate regime where $\Delta_N <\Delta_{\rm c} <\Delta_1$, which is just the aforementioned specific region in the protein-superconductor system and will be discussed in detail as follows. In this specific region, the protein could be divided into two segments at the specific site $N'$ so that $\Delta_ {N'+1} < \Delta_ {\rm c}< \Delta_{N'}$. In the left segment, all the pairing potentials satisfies $\Delta_{N'} >\Delta_{\rm c}$ and thus the zero modes are topologically trivial [see the red region in Fig.~\ref{fig:six}(c)], which is identical to Fig.~\ref{fig:six}(d). Contrarily, all the pairing potentials in the right segment satisfies $\Delta_ {N'+1} < \Delta_{\rm c}$ and the topologically nontrivial zero modes emerge [see the yellow region in Fig.~\ref{fig:six}(c)], just the same as Fig.~\ref{fig:six}(b). As a result, one zero mode will always be localized around the site $N'$, i.e., the topologically nontrivial-trivial interface [see the green-solid line in Fig.~\ref{fig:six}(c)]. This zero mode will be shifted from the left end to the right one by increasing $\Delta$, as can be seen from the white-oblique trajectories in Figs.~\ref{fig:five}(d) and~\ref{fig:five}(e). Although both topologically nontrivial and trivial zero modes exist in the protein-superconductor system, the overall topological property is mainly determined by the segment which dominates the electron transport along the protein. Specifically, when the topologically nontrivial range is longer than the trivial one, the protein-superconductor is topologically nontrivial with $Z_2=1$. Contrarily, when the topologically nontrivial range is shorter than the trivial one, the system is expected to be topologically trivial with $Z_2=0$. Therefore, the critical pairing potential $\Delta_\lambda$ of the topological phase transition for the decaying superconductivity is expressed as $\Delta_ {(N+1)/2}= \Delta_\lambda e^{-(N+1)/2 \lambda} = \Delta_{\rm c} $, which is just Eq.~(\ref{eq14}) derived from the numerical calculations.

\section{\label{sec4}Conclusions}

In summary, we investigate theoretically the topological properties of a single-helical protein sandwiched between a superconducting electrode and a normal-metal one. In this protein-superconductor system, the pairing potential attenuates exponentially along the helix axis of the protein, which is closely related to previous experiments. Our numerical results are qualitatively consistent with these experiments that a zero-bias conductance peak of conductance quantum is observed at zero temperature and the peak height (width) decreases (broadens) with increasing the temperature, and this zero-bias peak splits into two peaks by increasing the Zeeman field. Besides, this protein-superconductor system exhibits Majorana zero modes in a wide range of model parameters. In sharp contrast to homogeneous superconductivity, a specific region is demonstrated in the protein-superconductor system with decaying superconductivity, where the bandgap remains constant. In particular, both topologically nontrivial and trivial zero modes exist in this specific region, and the topologically nontrivial zero modes can transform to the trivial ones in the absence of bandgap closing-reopening. For the Majorana zero modes, the wave-function is localized at the left half part and the right end of the single-helical protein; whereas for the topologically trivial zero modes, it is localized at the right half part and the right end, which may provide an experimentally accessible way to discern the topologically nontrivial and trivial zero modes in the protein-superconductor system. In other words, instead of localized at the ends, one of the two zero modes could be localized at any position of the protein and be shifted from the left end toward the right one by increasing the pairing potential. Our theoretical work offers a unique perspective for the emergence of zero-bias conductance peaks in previous experiments and a deep understanding of the intriguing topological properties of chiral molecule-superconductor systems.

\section*{Acknowledgments}

This work is supported by the National Natural Science Foundation of China (Grants No. 11874428, No. 11874187, and No. 11921005), the National Key Research and Development Program of China (Grant No. 2017YFA0303301), and the High Performance Computing Center of Central South University.

\end{document}